\documentclass[12.5pt,a4paper]{article}
\usepackage{jheppub}

\pdfoutput=1 

\usepackage{hyperref}
\usepackage{amsmath}
\usepackage{graphicx}
\usepackage{slashed}
\usepackage{dcolumn}
\usepackage{bm}
\usepackage{multirow}
\usepackage{subfigure}

\title{Next-to-Leading Order Simulation of Slepton Pair Production}

\author{Il\'an Fridman-Rojas}
\emailAdd{ilan.fridman-rojas@durham.ac.uk}
\author{, Peter Richardson}
\emailAdd{peter.richardson@durham.ac.uk}
\affiliation{Institute for Particle Physics Phenomenology (IPPP), University of Durham \\
South Rd, Durham, DH1 3LE, UK.}

\abstract{The next-to-leading order (NLO) corrections to Drell-Yan slepton pair production are merged with parton shower evolution via the \textsf{POWHEG} method in \textsf{Herwig++}. The NLO corrections at the level of event generation alter the shape of observables in a non-global way, thereby potentially affecting the fraction of signal events which pass crucial transverse momentum ($p_T$) cuts on decay products and missing transverse momentum ($\slashed{p}_T$) requirements on events. This effect is not captured by an overall rescaling of the cross section and can have repercussions on the reach of LHC searches.}

\begin{document}
\vspace{25pt}

\begin{flushright}
IPPP/12/55\\
DCPT/12/110\\
MCNET/12/11
\end{flushright}
\maketitle

\newpage

\section{Introduction}

With LHC integrated luminosities well into ranges which allow for the search of potential new physics, the search for Supersymmetry (SUSY) is well and truly underway. As a model which, provided it is softly broken, offers solutions to the problem of the sensitivity of the Higgs mass to radiative corrections, can naturally produce electroweak symmetry breaking radiatively, produces unification of gauge couplings at the percent level, and provides a stable neutral dark matter candidate, SUSY is an extremely well motivated candidate for Beyond the Standard Model (BSM) physics.
 
As the combined data set taken at $\surd s =$ 7 and 8 TeV now has an integrated luminosity $\mathcal{O}(10\; \mathrm{fb}^{-1})$, the prospect of definitively confirming or conversely excluding broken supersymmetry at the weak scale is very much alive.

Furthermore given that limits from the currently ongoing searches are pushing the exclusion of colour charged sparticles towards masses $\mathcal{O}(1 \; \mathrm{TeV})$ in both the CMSSM and simplified models~\cite{CMS-PAS-SUS-12-002, ATLAS-CONF-2012-033} searches in the electroweak sector are increasingly relevant.

To conduct these searches with any confidence, potential signals and backgrounds must be modelled with the highest possible accuracy, minimising theoretical uncertainties (namely renormalization/factorization scale dependence).

If SUSY is discovered it will be important to determine which of the many UV-generated guises of SUSY and models created to provide a mechanism for SUSY-breaking (see~\cite{Lodone:2012kp} for a recent non-exhaustive overview) is realised in nature. Studies of how well we could deduce the point in SUSY parameter space which could have generated a given signal (the so called {\it inverse problem}) has been studied in Ref.~\cite{ArkaniHamed:2005px}, and then improved on in Ref.~\cite{Bornhauser:2012iy}. For example, this last study used leading-order (LO) matrix elements to generate the signal, and it is arguable that even for counting observables (rather than kinematic distributions), next-to-leading-order (NLO) corrections could produce sizable corrections as they could impact on acceptances after cuts.

Monte Carlo event generators presently available to produce predictions to test for generic SUSY spectra lag behind the progress that has been made in producing higher accuracy predictions for Standard Model (SM) predictions, and largely remain at the level of LO matrix elements rescaled by global NLO K factors (and merged with a parton shower (PS) which performs Leading Logarithmic (LL) all-orders resummation of soft and collinear logarithms).\footnote{NNLO SUSY-QCD corrections for non-coloured final states are expected to be very subdominant. Indeed already the NLO SUSY-QCD corrections to Drell-Yan dilepton + jet have been found to be very modest~\cite{Gavin:2011wn}. In terms of logarithmic contributions, the analytic resummation of $p_T$ and threshold logarithms for slepton pair production is known at next-to-leading-logarithmic (NLL) accuracy.~\cite{Bozzi:2007tea}}

The present work aims to begin to bridge this gap\footnote{Recently the implementation of LO multi-parton SUSY matrix elements matched with parton showers has begun and been performed for gluino/squark production~\cite{Dreiner:2012gx} and gaugino pair production~\cite{Fuks:2012qx}.} by implementing the known NLO corrections to a benchmark SUSY process into a Monte Carlo event generator via the Positive Weight Hardest Emission Generator \textsf{POWHEG} method~\cite{Nason:2004rx}, thus providing event generation with the $\mathcal{O}(\alpha_S)$ corrections to the matrix element used for simulation of production of slepton pairs via the Drell-Yan process.

Non-global changes to shapes of observables like $p_T$ can be crucial for signals which are expected to consist of only a handful of events. NLO corrections can either increase sensitivity to these events, or reduce it (relative to the LO prediction), depending generally on how heavy the relevant sparticles are, as well as potentially impact on the choice of optimal cuts for signal regions.

Production channels for slepton pairs (with the corresponding radiative corrections) are reviewed in section~\ref{pairchannels}, the merging of the NLO corrections and the parton shower are discussed in section~\ref{powhegsec}, and results using simplified SUSY models are presented in section~\ref{results}.


\section{Slepton Pair Production}
\label{pairchannels}

Slepton pairs can be generated in one of two ways: in decay chains from squark/gluino or gaugino pairs; or via direct pair production. We will discuss each one of these in turn.

Judging solely by production cross section, the first channel via which one could expect to have sensitivity to slepton production at a hadron collider is via squark/gluino pair production. These would promptly decay to coloured Standard Model (SM) particles and gauginos, which could themselves decay via sleptons. Alternatively, with smaller cross section but via potentially shorter decay chains, gaugino pair production could yield sleptons along their decay cascade (CMS has begun to study this scenario~\cite{CMSdilepton}).

However, in either of those scenarios where sleptons are a step in a decay chain, despite the possibly large original production cross section, the numerous stages in the decay imply suppression of the desired final state cross section by a product of several (potentially small) branching fractions. These branching fractions themselves also introduce extra model dependence as they depend sensitively on mass splittings and on mixings present in the gauginos involved in the decays.

The same is true even for direct slepton pair production if we allow for spectra where for example $m_{\tilde{l}} > m_{\chi_2^0}$ such that more than one decay can occur before yielding an observable final state.
Since we intend to stay as model-independent as possible we will therefore not consider any of the above scenarios. We restrict ourselves to direct slepton pair production.

The possible channels to pair-produce sleptons are Drell-Yan, vector boson fusion (VBF) and gluon fusion. However, at the LHC gluon fusion has been found to be negligible\footnote{Albeit it at a $\surd s = 16 \; \mathrm{TeV}$ LHC, though the conclusion should still hold and perhaps even be stronger given that for higher centre-of-mass energies one probes lower partonic momentum fractions where the gluon parton distribution functions dominate and are hence more likely to be the dominant channel.}~\cite{delAguila:1990yw}, and VBF has been argued to be subdominant for slepton masses up to $\approx 300 \; \mathrm{GeV}$, at which point it can become competitive\footnote{The NLO QCD corrections to this channel are known~\cite{Konar:2006qx} and are very modest, so higher order corrections are unlikely to make this channel competitive at low slepton masses.}~\cite{PhysRevD.37.2533, Choudhury:2003hq}. However, it is precisely around this mass where previous studies have found the expected reach of the LHC to lie~\cite{delAguila:1990yw, Baer:1993ew}, hence we will focus on Drell-Yan pair production only.

It is perhaps worth noting that working within the CMSSM at an 8 TeV LHC pair production of sleptons via Drell-Yan is sensitive mostly at very small $m_0$ and $m_{1/2}$~\cite{Kramer:2012bx}. However, in a framework more agnostic to the SUSY breaking scheme, such as the pMSSM or a simplified model, slepton final states provide independent channels to search for SUSY, as well as independent access to the electroweak sector of supersymmetric spectra.
Indeed it is these simplified models which we will be concerned with as a proof of concept in section~\ref{results}. A more thorough study of phenomenology will follow in an upcoming work.



\begin{figure}
\begin{center}
\includegraphics[scale=0.4]{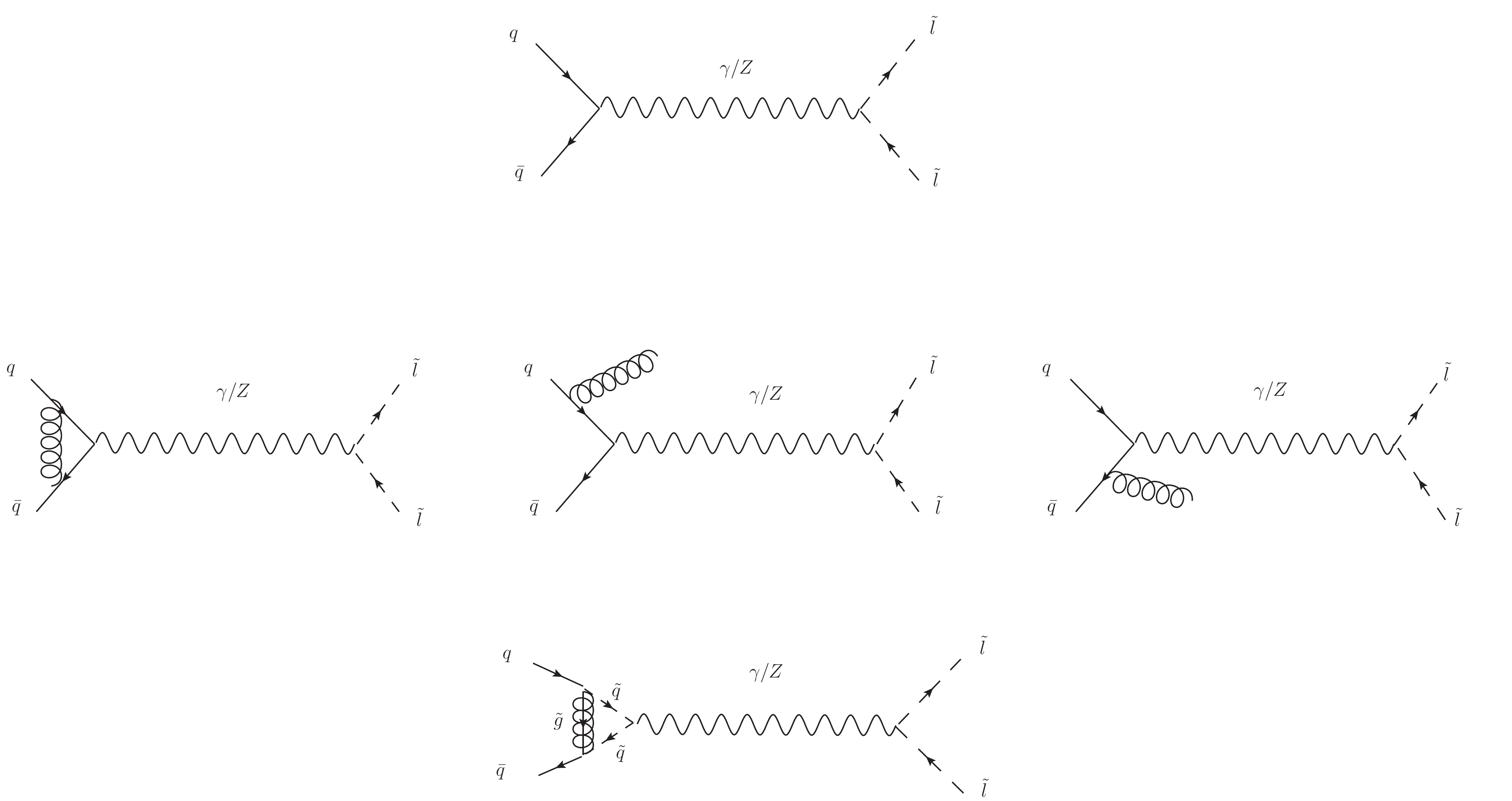}
\caption{The LO (first row), NLO QCD (second row) and NLO SQCD (bottom row) corrections relevant to slepton pair production (the diagrams for $W^{\pm}$ mediated processes are analogous). Self-energy corrections to the incoming partons not shown.\label{NLOdiag}}

\end{center}
\end{figure}

The LO cross section for Drell-Yan slepton pair production has been known for nearly 30 years~\cite{Eichten:1984eu}. The NLO QCD corrections to slepton pair production were calculated in Ref.~\cite{Baer:1997nh}, and the corresponding SUSY-QCD contribution followed in Ref.~\cite{Beenakker:1999xh}. These are implemented in the \textsf{PROSPINO2}~\cite{Beenakker:1996ed} package which computes fully inclusive total cross sections for SUSY processes at NLO in SUSY-QCD. It is these $\mathcal{O}(\alpha^2 \, \alpha_S$) diagrams which we are interested in, and which are shown in Figure~\ref{NLOdiag}. \\

\section{The POWHEG Method}
\label{powhegsec}

The motivations for incorporating NLO corrections as well as a parton shower are manyfold: the potential increase in cross section, the reduced scale uncertainty, the need to resum leading soft and collinear logarithms to all orders, the requirement to generate further jets and jet substructure reasonably faithfully (i.e. within the limits of the collinear approximation), and the need to more accurately estimate the fraction of signal events passing the imposed experimental cuts.

However, the merging of a NLO matrix element, which includes an emission in the real contribution, and a parton shower, which generates multiple emissions, raises the issue of double counting. That is, if care is not taken it can be possible for both the matrix element and the parton shower to radiate into the same region of phase space, effectively overestimating QCD radiation.

One of the two existing methods to consistently merge a NLO matrix element and a parton shower is given by the \textsf{POWHEG} method~\cite{Nason:2004rx}, and is the method chosen for this implementation. The corresponding
Standard Model Drell-Yan production of leptons has been available in this approach for some time~\cite{Alioli:2008gx,Hamilton:2008pd}.

The method is explained in detail in~\cite{Frixione:2007vw}, but schematically works as follows. The standard weight used to generate events in a Monte Carlo is given by\footnote{We omit most indices and show only relevant functional dependences for clarity. Collinear remnant terms are not shown but understood to be present.}
\begin{align}
\label{MC}
d\sigma \; &= \; B \; d\Phi_n \; \left[  \Delta(0) \; + \; \Delta(p_T) \; \mathcal{K} \; d\Phi_1 \right] \; , \\
	& \mathrm{where} \; \, \Delta(p_T^\mathrm{min}) = \exp \left( - \int \mathcal{K}  \; \Theta(p_T (\Phi_1) - p_T^\mathrm{min}) \; d\Phi_1 \right) \; \; . \nonumber
\end{align}
Here $d\sigma$ is the differential cross section of interest, $d\Phi_n$ is the $n$-body phase space relevant to the final state, $d\Phi_1$ is a single particle phase space, $B$ is the fixed order matrix element as will be defined below, $\mathcal{K}$ is a splitting kernel we will also define shortly, and the Heaviside step function acting on the transverse momenta imposes a cutoff to regulate the infrared singularities in the kernel.

Using this framework one can consistently merge a matrix element with a parton shower, producing results accurate at least to LO+LL. Progressive improvements can then be made from that starting point, to eventually produce NLO+(N)LL\footnote{For an appropriate choice of scale of the running coupling NLL accuracy can be reached as long as the process involves no more than 3 coloured external legs.~\cite{Frixione:2007vw}} results. The substitutions to equation~\ref{MC} required to achieve this are shown in Table~\ref{MCmod}.

\begin{table}[t]
\begin{center}
\begin{tabular}{|c|c|c|}

\hline
\hspace{5pt} & $B$ & $\mathcal{K}$ \\
\hline
LO+LL & $B$ & $P(z)$ \\
(MEC) LO+LL & $B$ & $\frac{R}{B}$\\
NLO+(N)LL & $\overline{B}$ & $\frac{R}{B}$ \\
\hline

\end{tabular}
\end{center}
\caption{The different choices of fixed order matrix elements and parton shower kernels used in equation~\ref{MC} to obtain the stated fixed order and logarithmic accuracy. Matrix Element Corrected (MEC) implies that a matrix element has been used to generate the hardest emission instead of the shower.}
\label{MCmod}
\end{table}

The notation in Table~\ref{MCmod} is as follows: $B$ is the Born amplitude squared\footnote{Factors of luminosity functions containing the relevant parton distribution functions are implied where relevant in all of the following.}, $P(z)$ denotes the unregularised DGLAP splitting function, $R$ is the real emission amplitude squared and $\overline B$ is the full NLO matrix element evaluated in the Born kinematics, defined as
\begin{align}
\overline{B} \, &= \, B \, + \, \left[ V \, + \, \int_1 \sum_\alpha C_\alpha \right]_{\epsilon=0} \, + \, \int_1 \, \left[R_{\epsilon=0} \, - \, \left(\sum_\alpha C_\alpha \, \right)_{\epsilon=0} \right] \; .
\end{align}
Here the integrals are over the single-particle phase space for the emission of an extra parton, $C_\alpha$ are dipole functions (in this case as defined in the Catani-Seymour dipole subtraction scheme~\cite{Catani:1996vz}, with the $\alpha$ subscript running over all possible choices of relevant dipoles), and the $\epsilon=0$ subscript denotes that the four dimensional limit of the dimensionally regularised matrix element is taken. 

To avoid double counting between the real emission matrix element and the parton shower, the exact form of $R$ which is used must be chosen carefully, both in the $\overline{B}$ term and the Sudakov form factor.

The alternative embodied by the \textsf{MC@NLO} method~\cite{Frixione:2002ik}  involves generating the shower as it normally would be done (i.e. using kernels in the soft and collinear approximation), but adding in the difference between the fixed-order real emission contribution and the parton shower approximation to it. This makes the implementation of this method dependent on the exact details of the parton shower algorithm it is being applied to, and involves a subtraction which can generate negative weight events which are unwieldy for detector simulations, which is why the \textsf{POWHEG} method was chosen for the present work.
However \textsf{MC@NLO} has been argued to have other advantages (for further discussion of the differences, strengths and weaknesses of the two approaches see~\cite{Hoeche:2011fd}). It is worth emphasizing that though some differences have been found between the two methods in certain observables, they are both equivalent up to higher order contributions, so the choice is a matter of preference.

The NLO+PS event generation thus proceeds by generating the hardest emission in the event according to the Sudakov form factor which uses $R$ in its kernel, then showering the event using the usual parton shower kernel (with the proviso that no further emissions be harder than the one generated according to $R/B$ occur).

Two further remarks are worth making: the NLO accuracy in the fully inclusive cross section derives from the fact that the terms in square brackets in equation~\ref{MC} can be shown to integrate to unity (this is what is commonly referred to as shower {\it unitarity}), thus yielding a cross section which integrates to $\int \overline{B} \; d\Phi_n$, which is the total NLO cross section.

Also since $\overline{B}$ is the NLO cross section (albeit in the Born kinematic limit), the weights produced by the \textsf{POWHEG} method are positive by construction, for as long as the process in question is perturbatively well-defined (specifically where the leading radiative corrections are considerably smaller than the LO contribution, and this itself is positive), $\overline{B}$ and hence weights produced from it should be positive.

For the present work we haven taken the virtual contribution from \textsf{PROSPINO2} and have generated the born, real emission and collinear remnant contributions independently. Factorisation and renormalisation scales were set equal to each other and set to the average mass of the outgoing sleptons, $\mu_{F/R}=\frac{1}{2}(m_3+m_4)$.


\section{Results}
\label{results}

\subsection{Validation}

Table~\ref{valtab} shows an example of a validation run against \textsf{PROSPINO2}. All cross sections are computed in the $\overline{\mathrm{MS}}$ scheme. Agreement at the sub per-mille level with \textsf{PROSPINO2} was found for all the SUSY points examined.

\begin{table}[htdp]
\begin{center}
\begin{tabular}{|c|c|c|c|c|}

\hline
\hspace{5pt} & \multicolumn{2}{|c|}{LO} &  \multicolumn{2}{|c|}{NLO} \\ \hline
Process  & \textsf{Herwig++} & \textsf{PROSPINO2}  & \textsf{Herwig++} & \textsf{PROSPINO2}  \\
\hline \hline

$\tilde{e}_L \; \tilde{e}_L$ & 4.1001(6) & 4.0982 & 4.811(2) & 4.8102 \\ \hline
$\tilde{e}_R \; \tilde{e}_R$ & 1.7646(3) & 1.7640 & 2.0788(9) & 2.0788 \\ \hline 
$\tilde{\tau}_1 \; \tilde{\tau}_2$ & 0.7956(1) & 0.79529 & 0.9314(4) & 0.93162 \\ \hline 
$\tilde{\nu}_e \; \tilde{\nu}_e$ & 4.0032(6) & 4.0020 & 4.688(3) & 4.6861 \\ \hline 
$\tilde{e}_L^- \; \tilde{\nu}_e$ & 4.2234(7) & 4.2230 & 5.084(2) & 5.0839\\ \hline 
$\tilde{e}_L^+ \; \tilde{\nu}_e$ & 11.407(2) & 11.401 & 13.039(6) & 13.042 \\ \hline 

\end{tabular}
\end{center}
\caption{Comparison of total cross sections for the SUSY point $m_0 = 500 \; \mathrm{GeV}$, \mbox{$m_{1/2} = 200 \; \mathrm{GeV}$}, $A_0 = 0 \; \mathrm{GeV}$, $\tan \beta = 10$, $\mu > 0$ at $\surd s \, = \, 14 \; \mathrm{GeV}$ at the LHC. All cross sections given in units of $10^{-1}$ fb. \textsf{Herwig++} errors are shown in brackets next to the digit they apply to, \textsf{PROSPINO2} errors are beyond the significant figures shown. Errors shown are numerical integration errors only.}
\label{valtab}

\end{table}%

Though we have implemented $W^{\pm}$ mediated Drell-Yan, we will not consider it for phenomenology purposes here because though it has the largest couplings and hence cross sections (see Table~\ref{valtab} for example), it produces fewer charged leptons and more missing transverse momentum. This implies it can be studied via less observables (namely $p_T^l$ and $m_T(l, \slashed{p}_T)$) and is more susceptible to larger SM backgrounds.

We will therefore focus on pair production of selectrons and smuons, focusing on selectrons for definiteness. We will further be conservative and consider right-handed selectrons as they have smaller couplings to the Z\footnote{In mSUGRA scenarios they will tend to be lighter than their left-handed analogues, but the difference in couplings more than compensates for the extra phase space.}, and for the simplified models which we will examine we require a single decay to be possible (namely $\tilde{e}_R \rightarrow \tilde{\chi}_1^0 \; e$), whereas the isospin of left handed sleptons could allow for decays via charginos if they are kinematically accessible.

Existing constraints on slepton masses come from direct searches from LEP~\cite{Heister2004247, opal, delphi, Achard200437} and ATLAS~\cite{ATLAS-CONF-2012-076}\footnote{CMS has so far only performed searches involving sleptons in decay chains but targeted at squark/gluino pairs and gaugino pairs~\cite{CMSdilepton}.}. Remaining as model-independent as possible, LEP roughly constrains $m_{\tilde{l}_{L/R}} \gtrsim 40 \; \mathrm{GeV}$ from bounds on the invisible width of the $Z$ boson (valid unless $m_{\tilde{l}_{L/R}} \approx m_{\tilde{\chi}_{1}^0}$), and likewise $m_{\tilde{\chi}_{1}^0} > 45.5 \; \mathrm{GeV}$~\cite{delphi}, assuming this neutralino eigenstate has a substantial wino component, otherwise there is no existing lower bound on its mass. If one chooses to subscribe to the CMSSM and assume gaugino mass unification, the overall LEP bound on sleptons then becomes roughly $m_{\tilde{l}_{L/R}} \gtrsim 100 \; \mathrm{GeV}$. In addition for the smuon, for the production of which there are no $t$-channel diagrams in $e^+e^-$ collisions, the limit is $m_{\tilde{\mu}_{L/R}}\gtrsim94$\,GeV provided that the mass difference between the smuon and lightest neutralino is $\gtrsim4$\,GeV~\cite{opal}.

\begin{figure}
\centering

\subfigure[The slepton pair invariant mass.]{\includegraphics[scale=0.62]{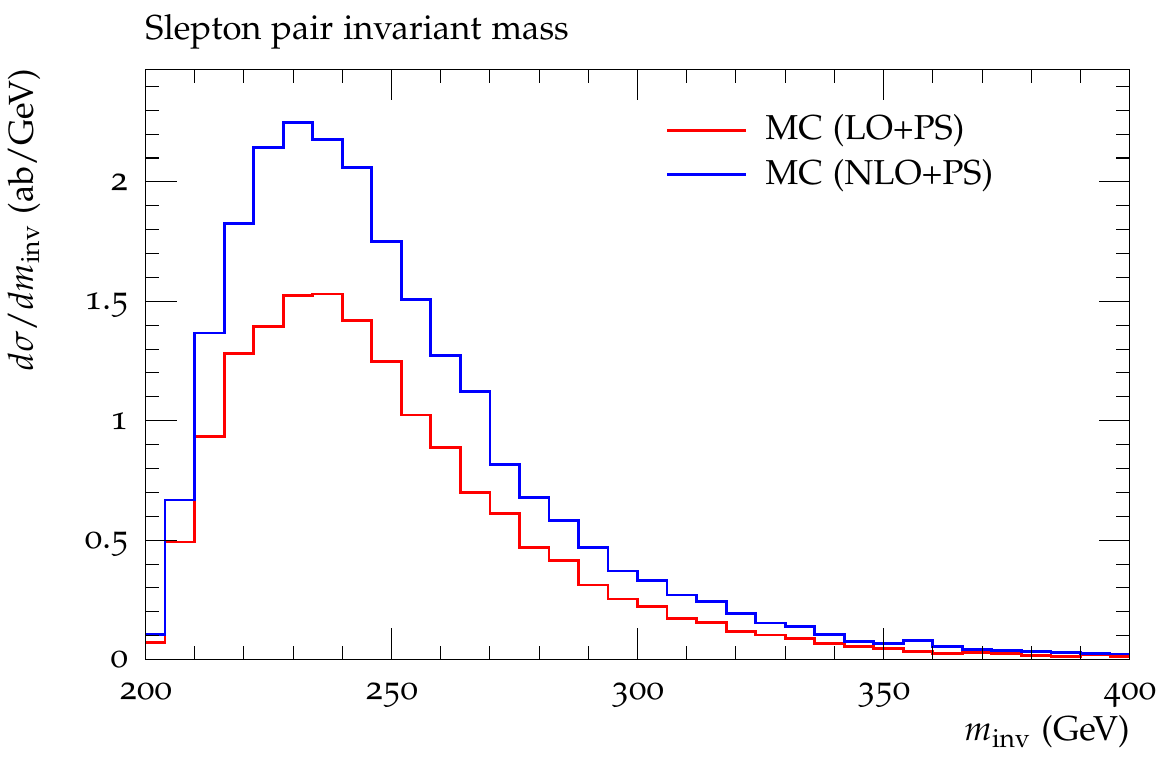}}
\subfigure[The slepton pair transverse momentum.]{\includegraphics[scale=0.62]{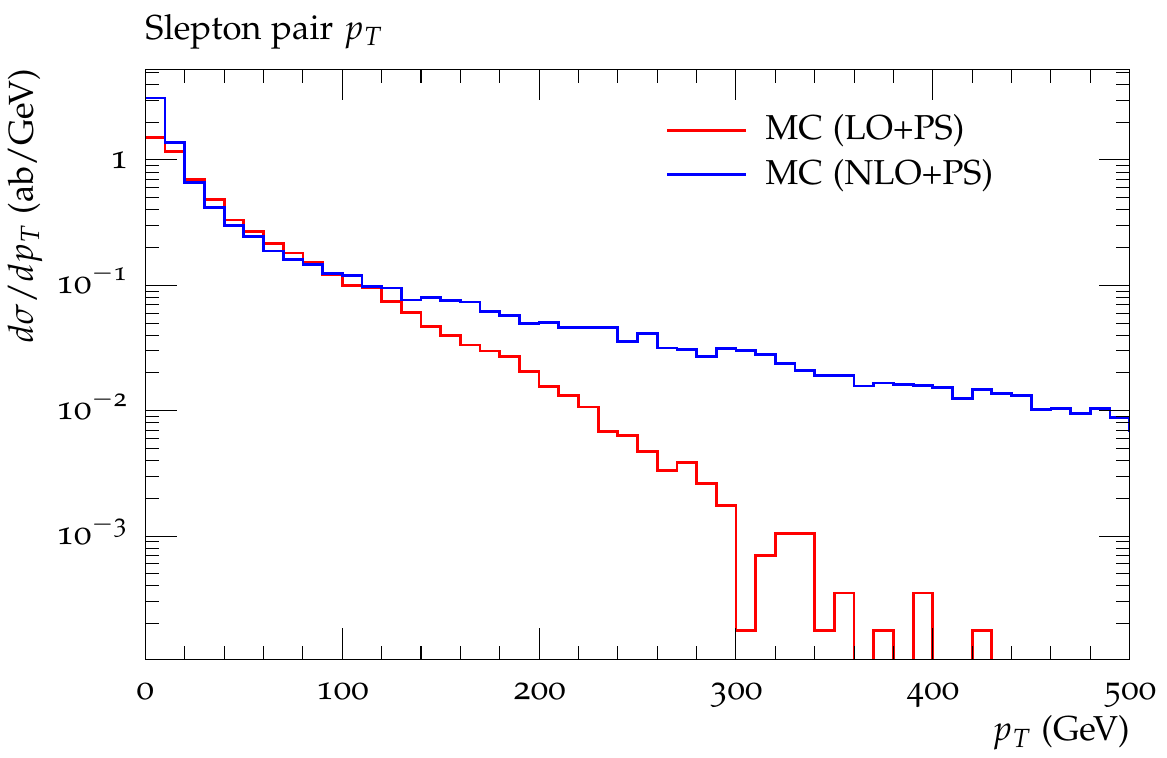}} \\

\subfigure[Missing transverse momentum.]{\includegraphics[scale=0.62]{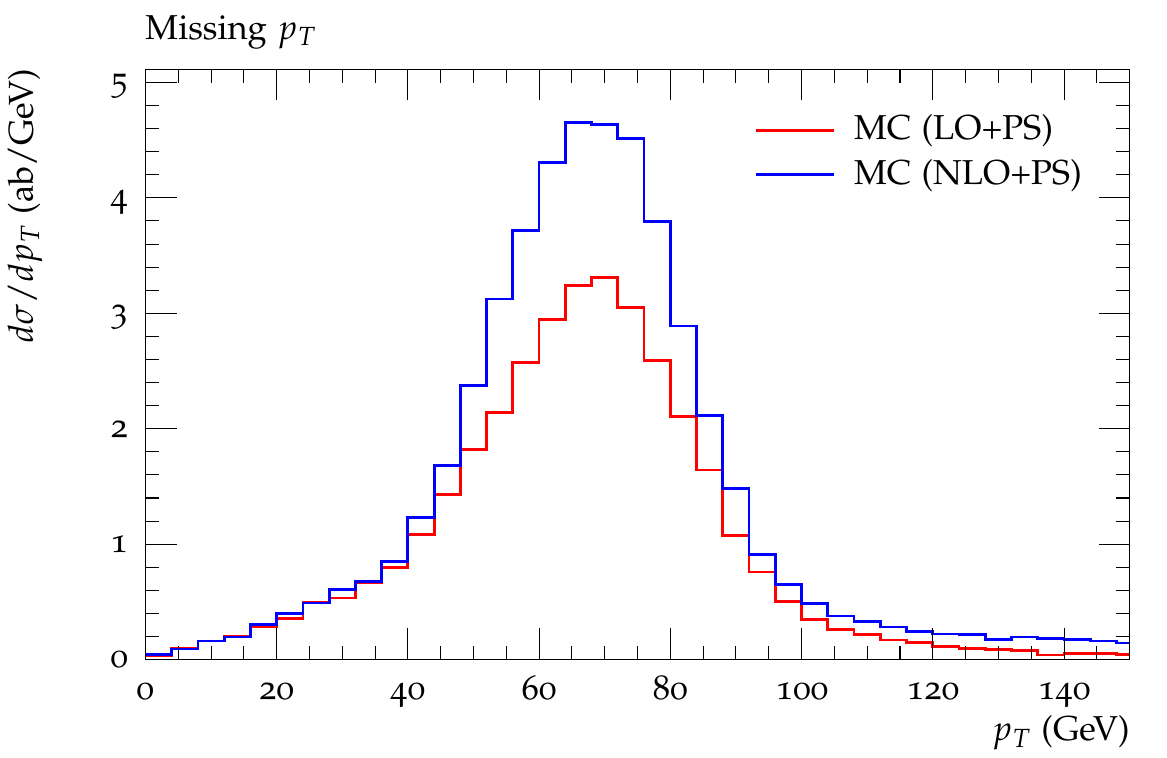}}
\subfigure[$m_{T2}$]{\includegraphics[scale=0.62]{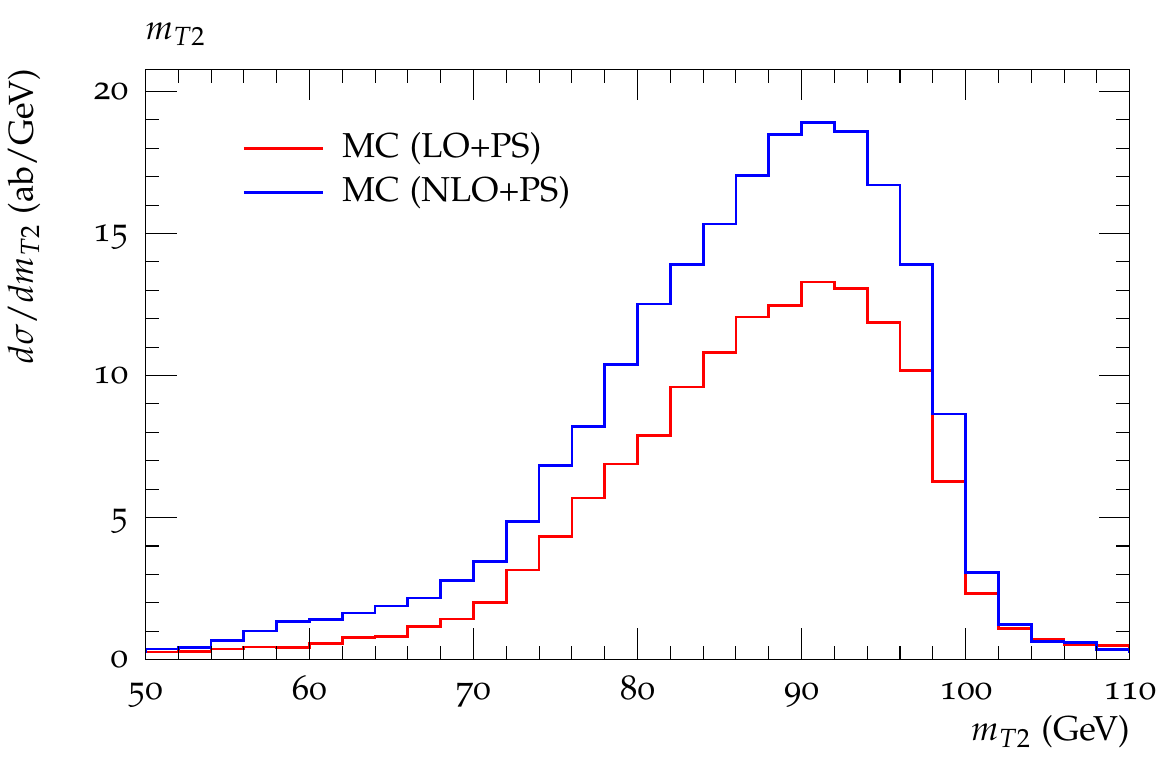}} \\

\subfigure[The selectron transverse momentum.]{\includegraphics[scale=0.62]{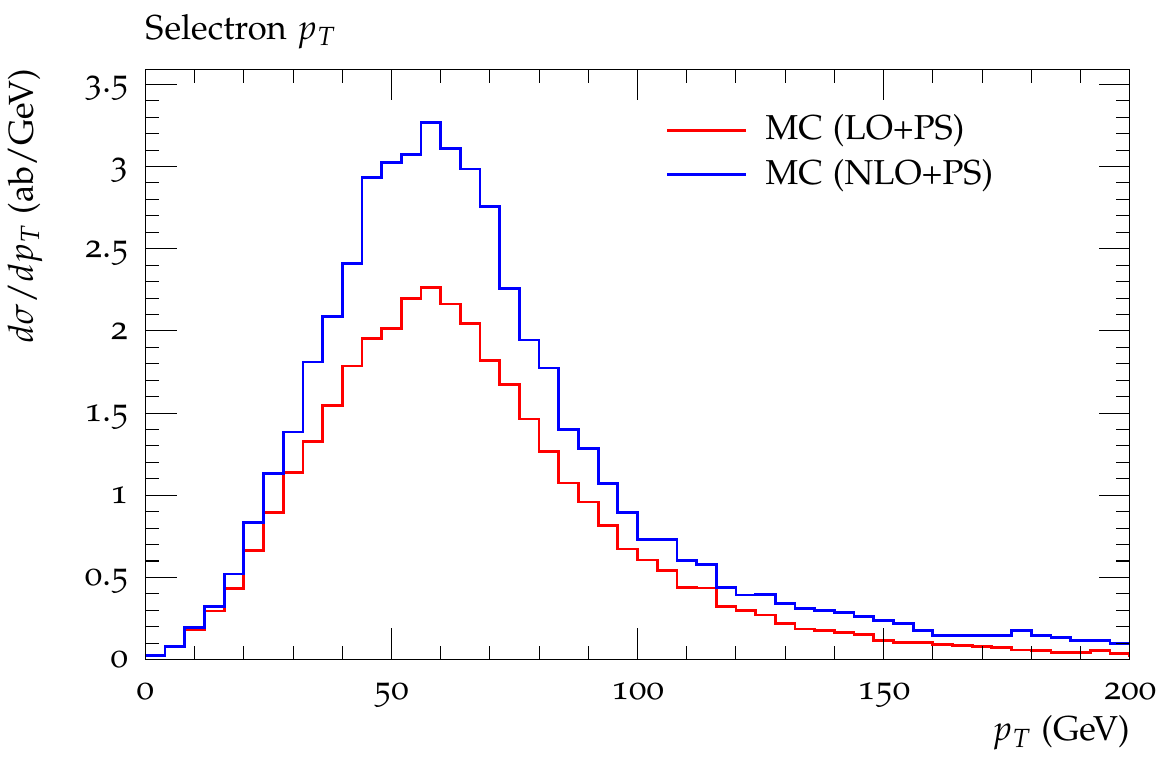}}
\subfigure[The electron transverse momentum.]{\includegraphics[scale=0.62]{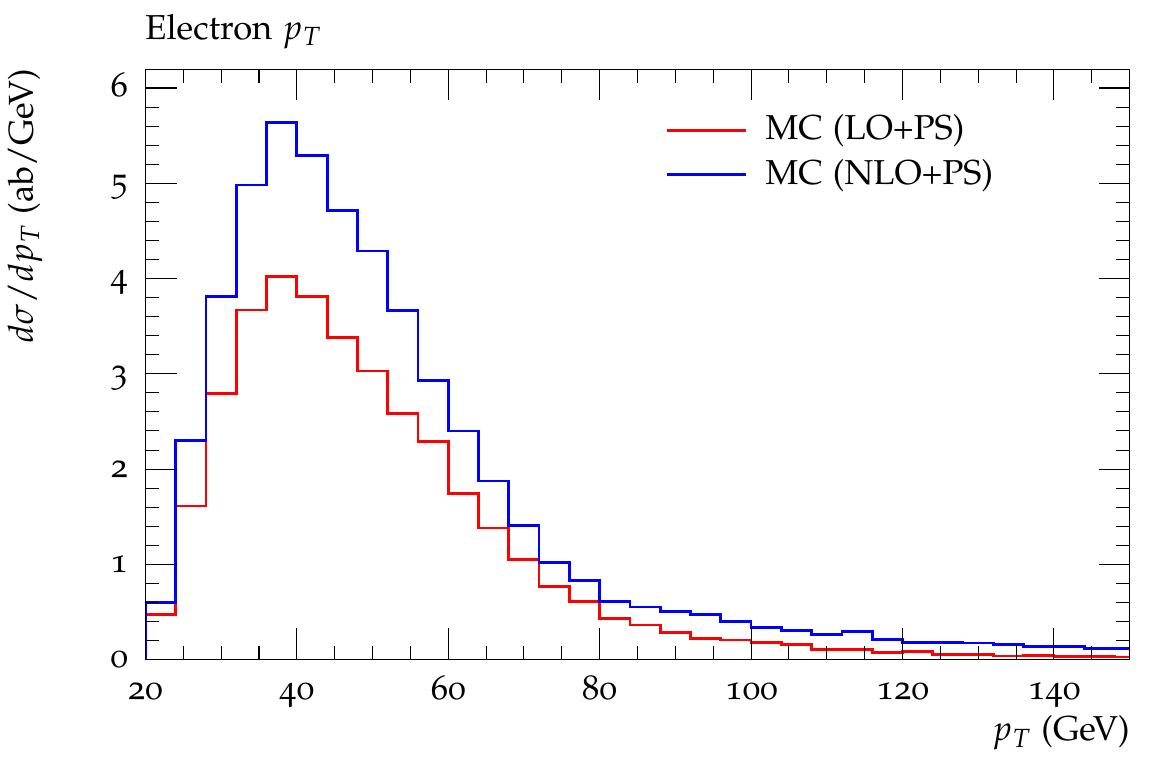}} \\

\caption{Comparison of observables produced by LO and NLO event generation for $\tilde{e}_R$ pair production in a simplified model with $m_{\tilde{e}_R} = 100 \; \mathrm{GeV}$, $m_{\tilde{\chi}_1^0} = 46 \; \mathrm{GeV}$  and $\mathrm{BR}(\tilde{e}_R \rightarrow \tilde{\chi}_1^0 \; e) = 1$ at $\surd s \, = \, 14 \; \mathrm{GeV}$ at the LHC. The total cross section for this point is $\sigma_{\mathrm{LO}} = 87.93(1) \; \mathrm{fb}$ and $\sigma_{\mathrm{NLO}} = 115.66(2) \; \mathrm{fb}$ with a K factor of $1.32$.}
\label{100_46_plots}
\end{figure}

\begin{figure}
\centering

\subfigure[The slepton pair invariant mass.]{\includegraphics[scale=0.62]{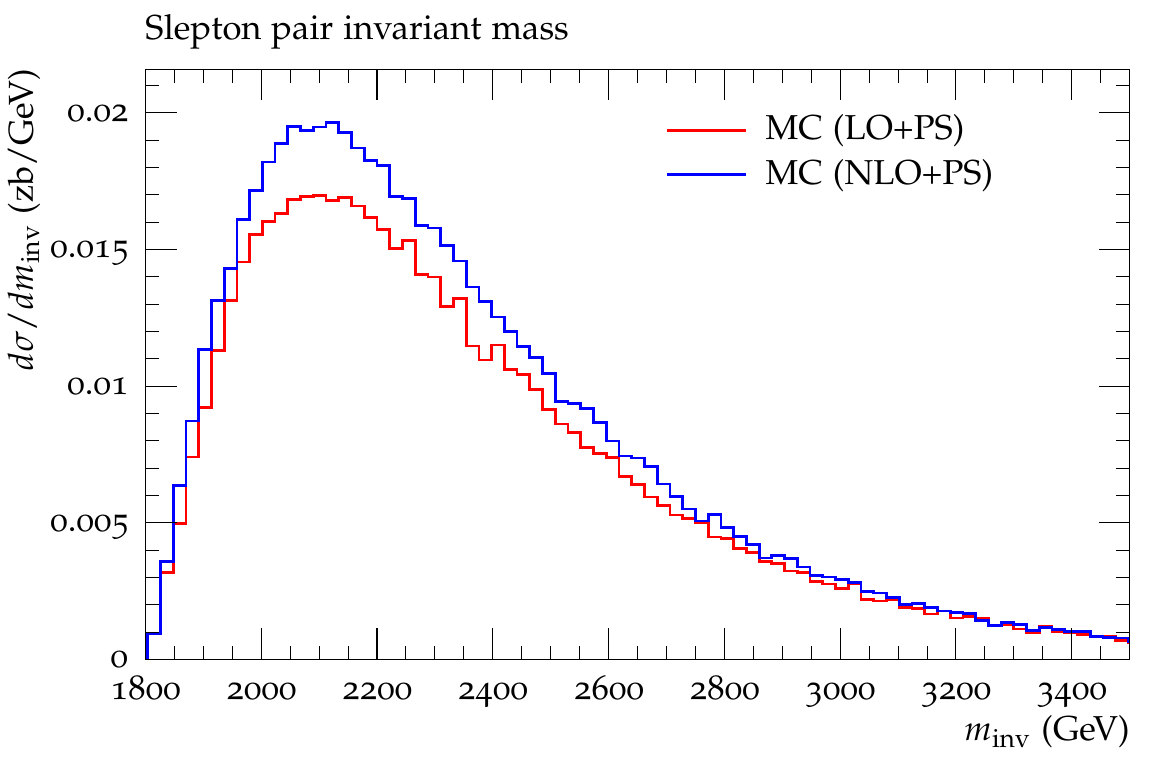}}
\subfigure[The slepton pair transverse momentum.]{\includegraphics[scale=0.62]{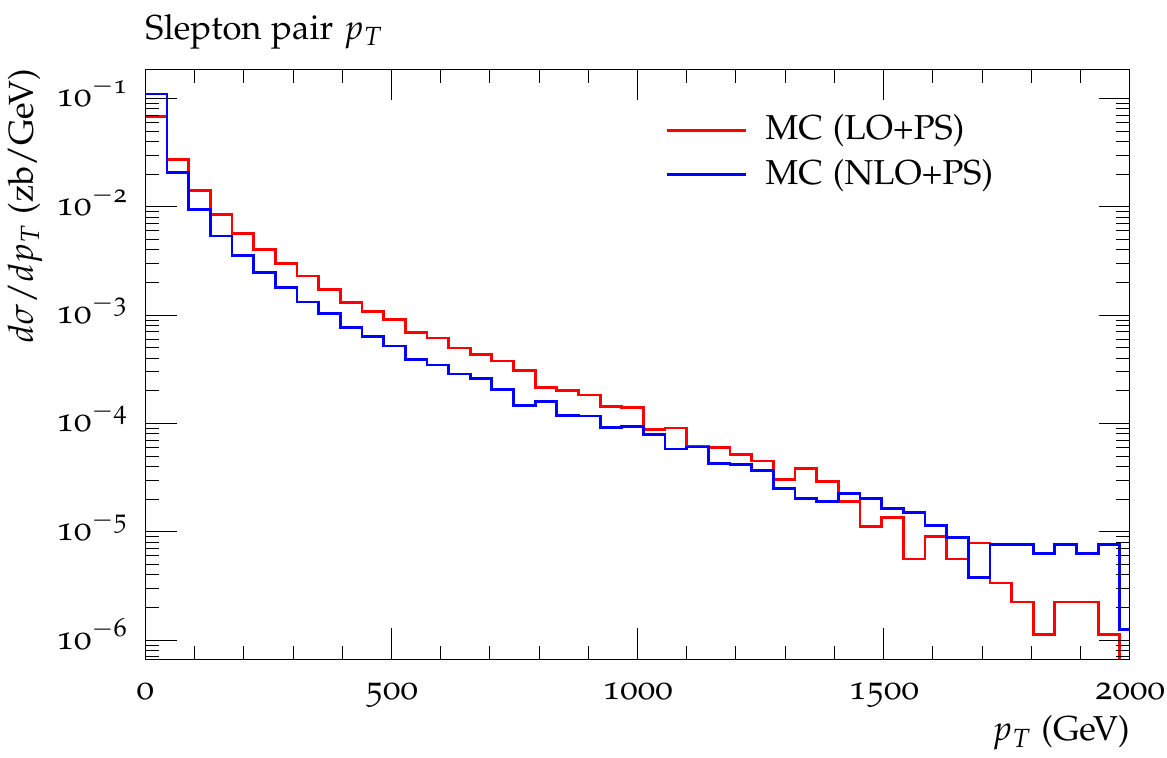}} \\

\subfigure[Missing transverse momentum.]{\includegraphics[scale=0.62]{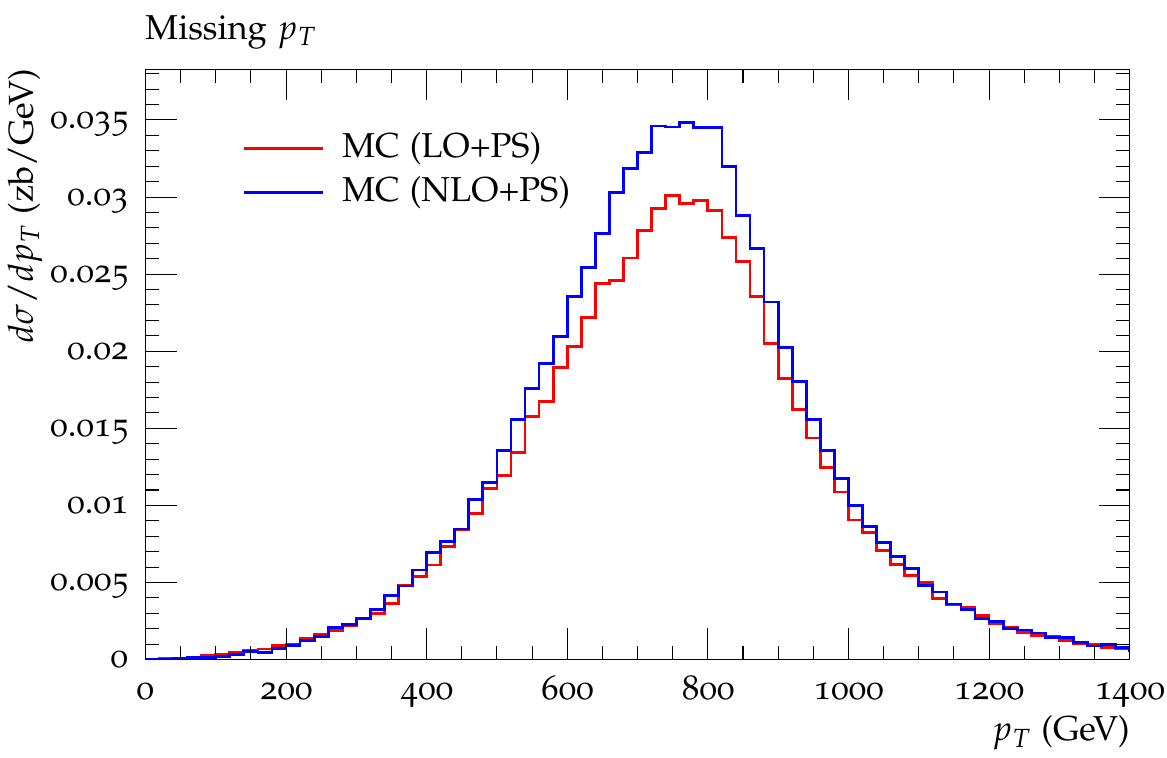}}
\subfigure[$m_{T2}$]{\includegraphics[scale=0.62]{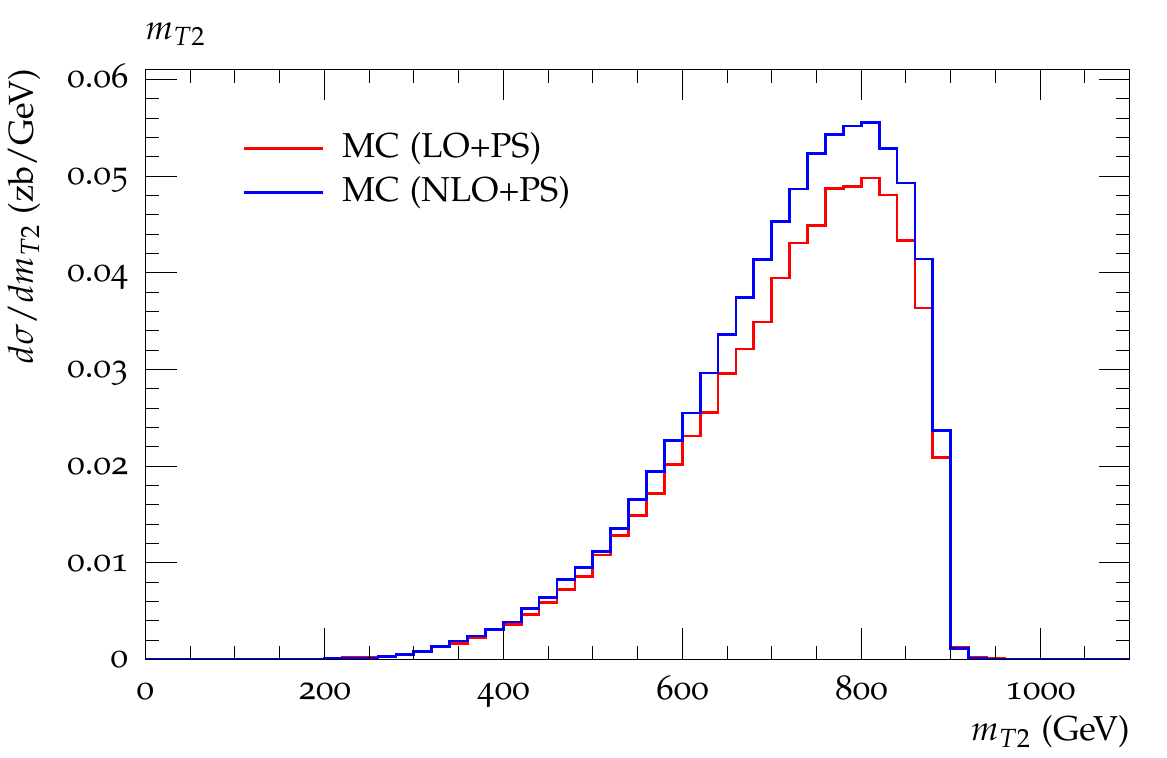}} \\

\subfigure[The selectron transverse momentum.]{\includegraphics[scale=0.62]{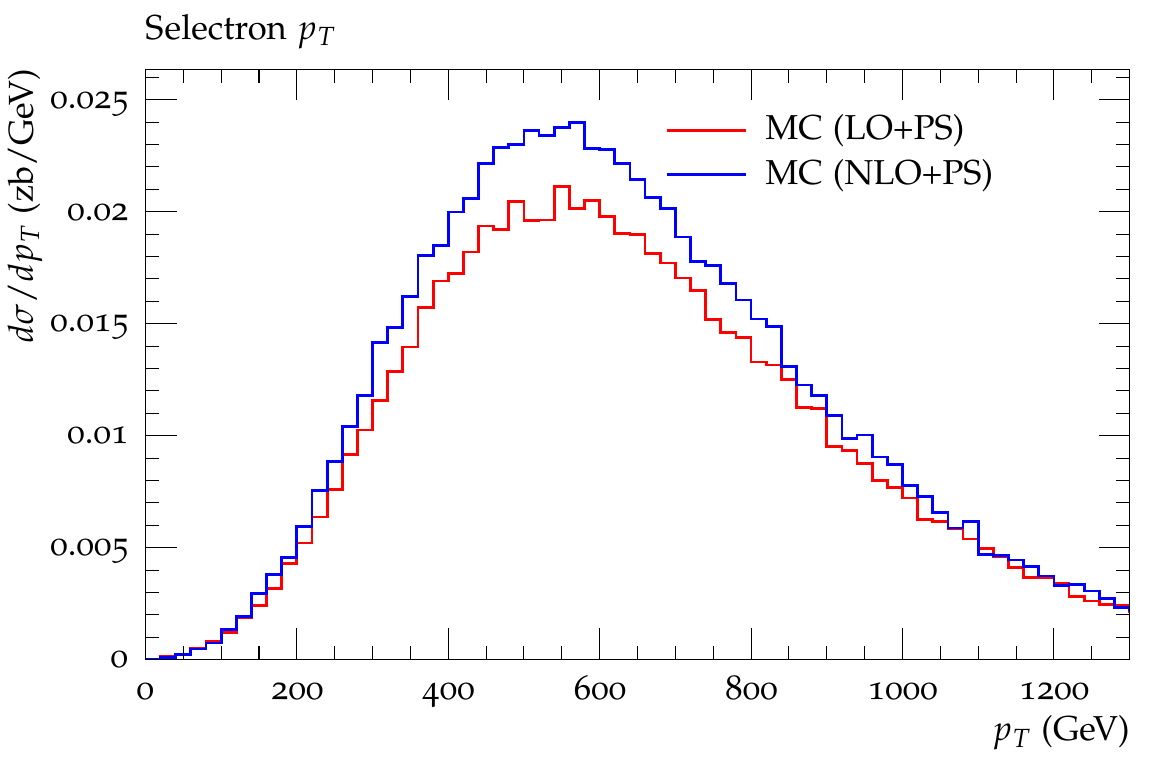}}
\subfigure[The electron transverse momentum.]{\includegraphics[scale=0.62]{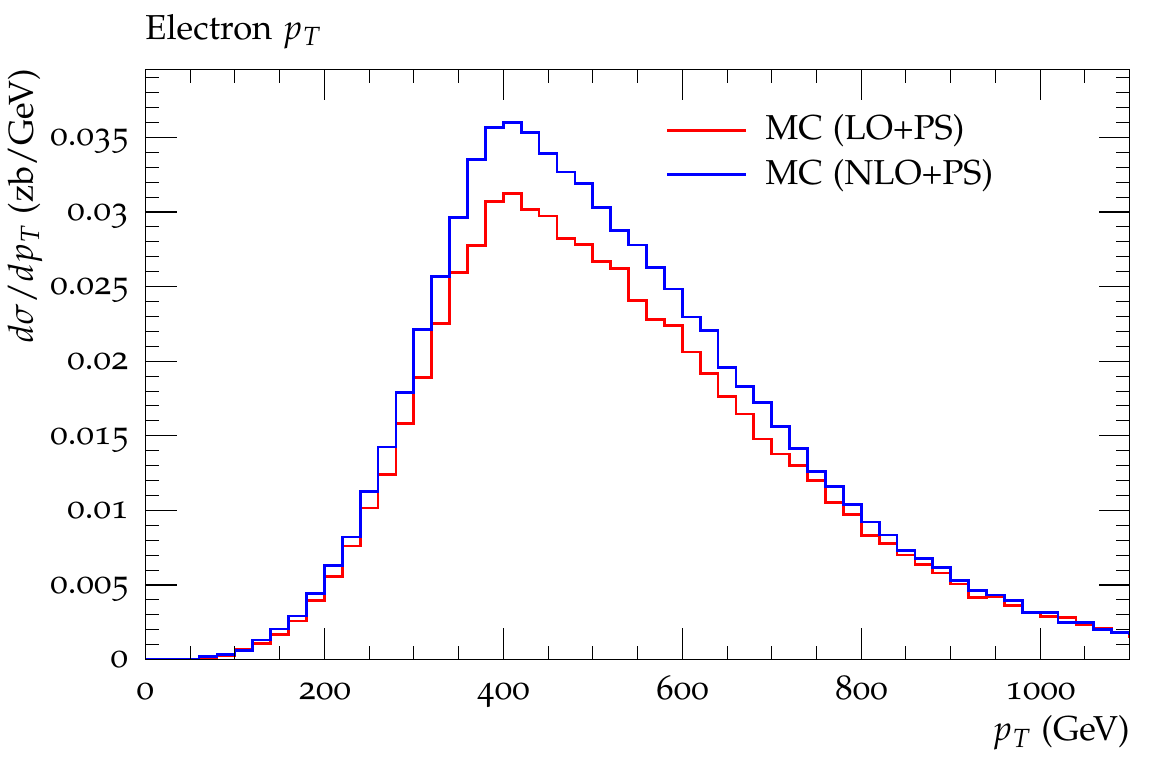}} \\

\caption{Comparison of observables produced by LO and NLO event generation for $\tilde{e}_R$ pair production in a simplified model with $m_{\tilde{e}_R} = 900 \; \mathrm{GeV}$, $m_{\tilde{\chi}_1^0} = 200 \; \mathrm{GeV}$  and $\mathrm{BR}(\tilde{e}_R \rightarrow \tilde{\chi}_1^0 \; e) = 1$ at $\surd s \, = \, 14 \; \mathrm{GeV}$ at the LHC. The total cross section for this point is $\sigma_{\mathrm{LO}} = 10.863(2) \; \mathrm{ab}$ and $\sigma_{\mathrm{NLO}} =  12.215(2) \; \mathrm{ab}$ with a K factor of $1.12$.}
\label{900_200_plots}
\end{figure}

In keeping with these existing bounds we have chosen two simplified models to illustrate a spectrum sitting just outside the exclusion, see Figure~\ref{100_46_plots}, and one which is one of the most commonly found in still allowed spectra from the pMSSM~\cite{CahillRowley:2012cb}, see Figure~\ref{900_200_plots}. Relevant transverse momentum and invariant mass observables are shown, including the stranverse mass as defined in Ref.~\cite{Lester:1999tx}. For now we have not included any of the relevant backgrounds, whether SM ($t \bar t$, $WW$, $ZZ$, $WZ$) or SUSY ($\tilde{\chi}^0_1 \; \tilde{\chi}^0_2$).

%
%



For both of these plots leptons and jets within pseudorapidity $[-2.5,2.5]$ were considered, with a $p_T$ cut of $20 \; \mathrm{GeV}$ on the leptons, and jet clustering via the anti-$k_t$ algorithm with a cone size of 0.4. Leptons were required to have a separation of at least $\Delta R \geq 0.4$ from any jet, and at least two leptons were required per event. In both cases the trial mass used to compute $m_{T2}$ was also set exactly equal to $m_{\tilde{\chi}_1^0}$. These settings are only meant to provide an illustration of what effect the NLO corrections can have.

It is also perhaps worth noting that unlike when using LO matrix elements, at NLO the production cross section remains sensitive to the masses of fields running in the loops of the virtual correction (in this case squark and gluino masses). For the examples shown here we have set them to $\surd s /2 = 7 \; \mathrm{TeV}$. This choice makes them clearly kinematically inaccessible to decays and has little effect in the virtual corrections which 
tend to the normal QCD corrections in the decoupling limit.

From both of the examples shown one can see that NLO contributions are significant and affect observables in a non-global way. In particular they increase the number of signal events which would pass a $p_T$ cut on the leptons or a $\slashed p_T$ cut on the event in a non-trivial way, which will depend on the exact position of the cut. This difference may be negligible in some regions, but could be significant if the signal constitutes only a few events.

%

\section{Conclusions}
NLO SUSY-QCD corrections to Drell-Yan slepton pair production have been implemented in \textsf{Herwig++} by use of the \textsf{POWHEG} method. 
 Validation of the total cross section was performed against \textsf{PROSPINO2} and good agreement was found.

From initial results simulating signal only and without full use of the full event generation machinery (multiple parton interactions, underlying event, QED final state radiation, etc) it was found that the NLO corrections are sizable and significantly affect observables in regions where experimental cuts are typically imposed. For signal rates which are expected to be small the effect of this could be significant and could result sizable corrections (relative to the LO predictions) to the reach of experimental searches. Moreover the NLO corrections change the shape of observables and hence the effects a particular set of cuts on the number of signal events passing them cannot accurately be accounted for by the usual rescaling by a an overall K factor. This could produce significant differences in signal samples that may consist of only a few events.

Given that the experimental collaborations at the LHC are beginning to reach integrated luminosities where they begin to become sensitive to the electroweak sector of SUSY models (and indeed they have already begun to produce results) we have shown that the use of NLO-accurate event generation could be important. Quantifying just how important would require a more detailed study of the effects of NLO corrections on LHC searches and will be undertaken in a future work.

This implementation can be used to study the effects of NLO corrections on studies which involve other observables not considered in the present work, such as those which might allow for determination of slepton spin~\cite{Barr:2005dz} or allow for good signal-background discrimination (such as the azimuthal angle between the leptons, as studied in~\cite{delAguila:1990yw}). It could also potentially extend the reach in searches for models with naturally light sleptons~\cite{Baer:2010rz}, or those in which sleptons are effectively stable.


\section*{Acknowledgements}
We acknowledge helpful discussions with Tilman Plehn and members of the \textsf{Herwig}++ collaboration.
IFR also acknowledges helpful discusions with Marek Sch\"onherr and Andrew Papanastasiou as well as financial support from CONACyT\@.

\bibliographystyle{JHEP}
\bibliography{slepton}

\end{document}